# Integrating PETs into Software Applications: A Game-Based Learning Approach

*Short Paper*

**Maisha Boteju**
University of Auckland
New Zealand
mbot450@aucklanduni.ac.nz

**Thilina Ranbaduge**
Data61, CSIRO
Canberra, Australia
thilina.ranbaduge@data61.csiro.au

**Dinusha Vatsalan**
Macquarie University
Australia
dinusha.vatsalan@mq.edu.au

**Nalin Arachchilage**
RMIT University
Australia
nalin.arachchilage@rmit.edu.au

**Abstract**

The absence of data protection measures in software applications leads to data breaches, threatening end-user privacy and causing instabilities in organisations that developed those software. Privacy Enhancing Technologies (PETs) emerge as promising safeguards against data breaches. PETs minimise threats to personal data while enabling software to extract valuable insights from them. However, software developers often lack the adequate knowledge and awareness to develop PETs integrated software. This issue is exacerbated by insufficient PETs related learning approaches customised for software developers. Therefore, we propose "PETs-101", a novel game-based learning framework that motivates developers to integrate PETs into software. By doing so, it aims to improve developers' privacy-preserving software development behaviour rather than simply delivering the learning content on PETs. In future, the proposed framework will be empirically investigated and used as a foundation for developing an educational gaming intervention that trains developers to put PETs into practice.

## Introduction

Privacy regulations, such as the General Data Protection Regulation (GDPR) (Commission 2016) and the public's heightened privacy awareness, compel software applications to safeguard personal data (Iwaya et al. 2023). Software applications depend on vast amounts of personal data for product personalisation and innovation (Hargitai et al. 2018; Sanderson et al. 2023). However, unprotected personal data can be leaked to unauthorised parties, causing financial losses, reputational damage, and emotional distress to the end-users (Boteju et al. 2024) (Senarath et al. 2019). For example, the Ashley Madison website, an extramarital affairs facilitator, was breached in 2015, leaking the names, contact details, locations, and financial transactions of 30 million users (Mansfield-Devine 2015). The breach resulted in damaged reputations, embarrassment, divorce cases, and suicides among the affected users (Mansfield-Devine 2015). In addition, software organisations also face financial losses and reputational risks from data breaches occurring through their software (Senarath and Arachchilage 2018b). For example, they might incur hefty fines imposed by privacy regulations (Sanderson et al. 2023). Thus, software applications require proactive data protection measures to mitigate data breaches and their consequences.

Software applications can achieve data protection by design and default through the integration of Privacy Enhancing Technologies (PETs) (Klymenko et al. 2022). PETs are technical measures that guarantee data protection by transforming data, hiding or shielding data, and splitting or controlling access to data (Office 2023). For instance, homomorphic encryption is a PET that allows computations on encrypted data without decrypting to its original form (data hiding) (Office 2023). Leveraging PETs for data protection is also supported by global standards (ISACA 2024; ISO 2018a). By utilising PETs, software can maintain the utility of personal data while minimising unauthorised identification and inference generation of end-users (Agrawal et al. 2021). Thus, PETs enable software to achieve data protection proactively, minimising privacy risks (Garrido et al. 2022).

Despite the data protection capabilities of PETs, they are not widely adopted by software developers (Agrawal et al. 2021). As discovered through our Systematic Literature Review (SLR) (Boteju et al. 2024), a prominent reason for this is the lack of PETs related knowledge among developers. Developers struggle to decide which PETs to use, when to apply them, and how to streamline the integration using support tools like libraries and APIs (Agrawal et al. 2021; Klymenko et al. 2022). These difficulties are aggravated by insufficient training and educational approaches to teach developers about PETs.

Many software organisations do not provide technical training sessions on data protection (Senarath and Arachchilage 2018a). Additionally, PETs related research and white papers contain complex information, which developers may find difficult to follow and understand (Agrawal et al. 2021). Even established privacy principles like Privacy by Design (PbD) (Cavoukian 2009) and privacy regulations lack guidance on integrating PETs into software (Senarath and Arachchilage 2018a). Thus, the absence of PETs related knowledge demotivates developers to utilise them for privacy-preserving software development, regardless of the efforts made by researchers to introduce or enhance PETs and provide support through libraries and APIs (Dwork et al. 2019). This realisation leads us to the following research question:

**RQ1: How can software developers be educated on PETs to improve their privacy-preserving software development behaviour?**

To address this research question, we aim to develop a game-based learning intervention for developers. We selected a game-based approach to enhance learner performance through an engaging and interactive learning experience (Alhazmi and Arachchilage 2023). However, to improve privacy-preserving development behaviour, developers need to put the learnt knowledge of PETs into practice. Thus, our learning intervention should be designed in a way that motivates developers to use PETs in software. To achieve this, we first propose a theoretical learning framework, "PETs-101", including the relationship between behaviour and motivation, factors generating motivation (motivational factors) and mechanisms to achieve those motivational factors in a game setting. This theoretical framework will guide the game intervention's design, implementation, and evaluation (Arazy et al. 2010). "PETs-101" was inspired by the findings of our SLR (Boteju et al. 2024), the Technology Threat Avoidance Theory (TTAT) (Liang and Xue 2010) and developer-focused privacy games (Alhazmi and Arachchilage 2023; Shilton et al. 2020). This paper discusses the theoretical underpinnings of "PETs-101" (Figure 1) with an example game scenario explaining its translation into a game setting. In future, "PETs-101" will be validated through a user study. The validated framework will be then used as a foundation to develop the game-based learning intervention.

## Background and Related Work

Software developers underutilise or misuse PETs as they lack knowledge of PETs (Boteju et al. 2024). Developers are unaware of various types of PETs, hindering their ability to select the most suitable ones for their applications (Hargitai et al. 2018; Senarath and Arachchilage 2018a). They often confuse PETs with security measures such as encryption (Hargitai et al. 2018). Some often resort to naive PETs like pseudonymisation (replacing direct personal identifiers with artificial identifiers) even when those techniques do not meet the expected data protection levels (Senarath and Arachchilage 2018b). For example, pseudonymised data are vulnerable to re-identification attacks when used in applications that handle extensive personal data; however, developers apply immature workarounds for such limitations (e.g., maintaining the secrecy of the integration process), perceiving a false satisfaction with the integration process (Hargitai et al. 2018).

In addition, developers struggle to understand the mathematical and theoretical concepts behind emerging PETs, such as differential privacy (injecting random noise into data), homomorphic encryption, and synthetic data generation (artificially generating data to mimic real data) (Hargitai et al. 2018; Sarathy et al. 2023). They fail to understand and calibrate parameters in PETs while balancing the privacy and utility trade-offs (Munilla Garrido et al. 2023). For example, some developers reuse parameter values used in research studies, overlooking the specifications of their applications in different contexts or problems, which then compromise data protection outcomes and other software goals (e.g., functionality) (Dwork et al. 2019). Further, it is challenging for developers to utilise the existing support tools due to their inability to comprehend the inner workings of PETs (Dwork et al. 2019; Sarathy et al. 2023).

Few studies have provided PETs related educational solutions for software developers. The "Epsilon Registry" is a knowledge-sharing approach where software organisations can share their decisions regarding differential privacy integration (Dwork et al. 2019). Additionally, the literature also suggests the need for readily available tutorials (Sarathy et al. 2023). However, these solutions are conceptual, lacking details and practical evaluation of concrete implementations. In contrast, the Privacy Knowledge Base (PKB) tool is an implemented educational approach that allows developers to identify appropriate PETs to mitigate privacy vulnerabilities (Baldassarre et al. 2021). However, it lacks theoretical and practical knowledge on implementing those PETs. Further, educational games focused on Privacy by Design (PbD) (Shilton et al. 2020) and GDPR (Alhazmi and Arachchilage 2023) provide an engaging way to teach developers about incorporating privacy into software. However, it is unclear what lessons or exercises they use to teach developers the technical measures (e.g., PETs) to achieve PbD or GDPR principles.

The above approaches fail to provide a progressive learning experience on PETs, i.e., from lower learning levels (e.g., understanding) to higher learning levels (e.g., applying) (Anderson and Krathwohl 2001). Further, they are not customised for developer expectations, such as providing technical examples. In contrast, "PETs-101" suggests a dedicated educational module with context-based lessons, exercises, and feedback mechanisms, which are the features that developers expect from a learning approach (Boteju et al. 2024). The knowledge delivered through "PETs-101" will convey how PETs are useful for data protection while improving developers' ability to use PETs. However, to actively use PETs, developers also require a mindset that prioritises privacy (Boteju et al. 2024). For example, developers who think end-users are solely responsible for protecting their privacy fail to address privacy concerns in software applications proactively (Boteju et al. 2024). Thus, besides providing knowledge, "PETs-101" attempts to achieve a perceptual change that motivates developers to use PETs.

## Framework Overview

As Figure 1 highlights, "PETs-101" intends to influence developers' behaviour in developing privacy-preserving software using PETs. For this, it attempts to motivate developers to utilise PETs. The components in "PETs-101" are derived from the TTAT (Liang and Xue 2010), Boteju et al. (2024) and developer-based privacy educational games (Alhazmi and Arachchilage 2023). From the TTAT, we identified the relationship between behaviour and motivation, along with some motivational factors. The TTAT discusses how Information Technology (IT) end-users are motivated to use threat-avoiding safeguards (Liang and Xue 2010). We modified the selected TTAT components to the developers' perspective. Those modifications are supported by the findings of Boteju et al. (2024). Finally, Alhazmi and Arachchilage (2023) and Shilton et al. (2020) inspired the mechanisms to achieve the identified motivational factors in a gaming intervention. These mechanisms are discussed using an example game application.

### *Privacy-Preserving Software Development Behaviour and Motivation*

The first step is to build a relationship between privacy-preserving development behaviour and motivation to use PETs. Motivation is the force that directs humans to achieve goals by performing required behaviours (Ajzen 1991). The TTAT depicts how IT users with threat avoidance motivation use safeguards (e.g., antivirus software) to perform threat avoidance behaviour (Liang and Xue 2010). Thus, we argue that developers' motivation to use PETs (avoidance motivation) leads them to develop PETs embedded software, thus achieving privacy through personal data protection (privacy-preserving behaviour).

H1: Motivation to integrate PETs into software positively influences developers' privacy-preserving development behaviour.

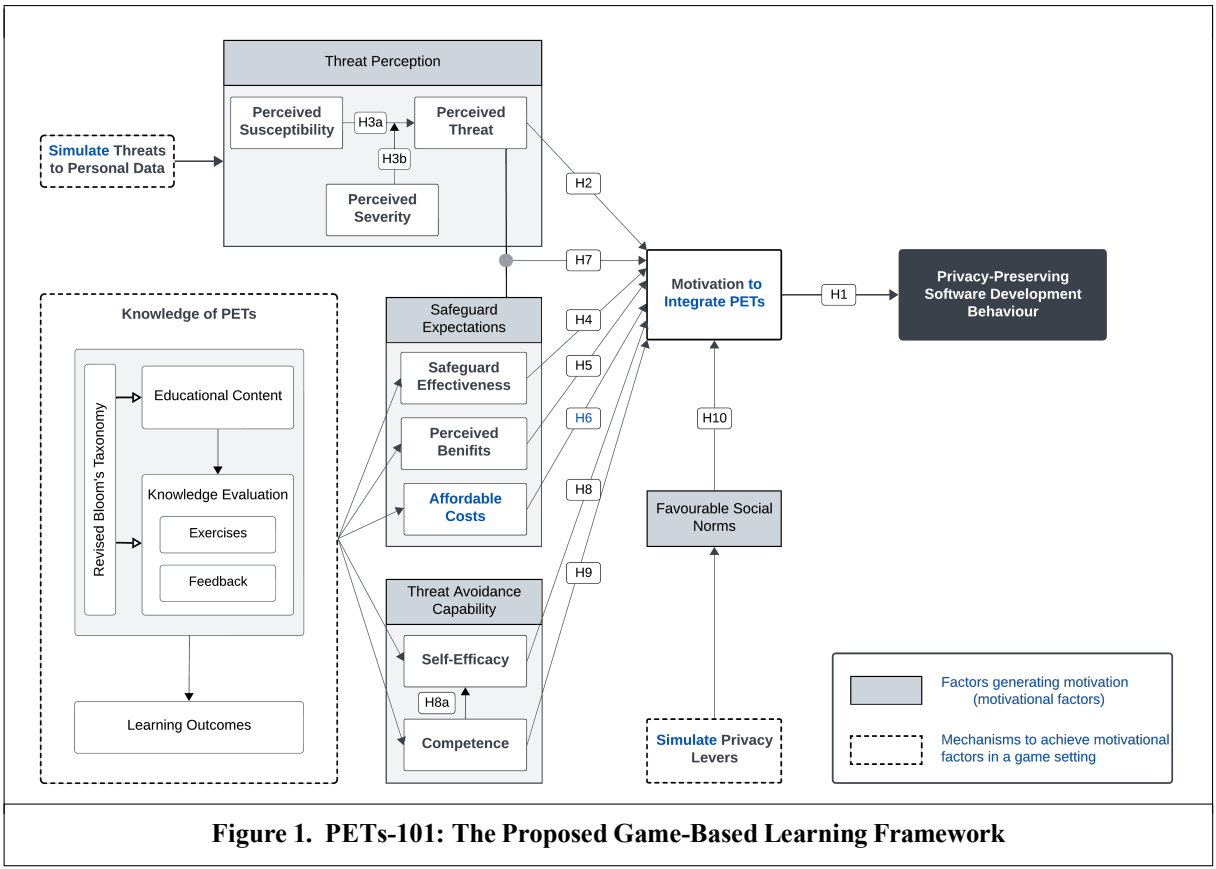

**Figure 1. PETs-101: The Proposed Game-Based Learning Framework**

## *Motivating Developers to Integrate PETs into Software*

"PETs-101" categorises the identified motivational factors into four components: threat perception, safeguard expectations, threat avoidance capability, and favourable social norms. First, developers should identify the potential data exploits within their software applications, i.e., Threat Perception, to identify the need for PETs. Second, PETs should meet certain developer expectations as safeguards (e.g., effectiveness and additional benefits), i.e., Safeguard Expectations. Third, developers' ability to mitigate the identified threats using certain PETs is considered, i.e., Threat Avoidance Capability. Finally, the Favourable Social Norms capture how others around developers favour PETs.

**Threat Perception**

People are inherently motivated to avoid negative experiences (Tanner et al. 1991). For example, health psychology has argued how people become more health conscious when they realise the threats to their health (Weinstein 1993). The TTAT also translates this relationship to discuss how IT users are motivated to use safeguards once they realise threats to their systems (Liang and Xue 2010). Similarly, developers are influenced to use PETs when they perceive threats to personal data in their applications (Boteju et al. 2024). For example, a developer is motivated to use pseudonymisation to mask personal data, perceiving a threat of unauthorised access (Hargitai et al. 2018). Thus, in this context, we hypothesise:

H2: Perceived threat positively affects the motivation to integrate PETs into software.

Further, two antecedents support perceiving threats: perceived susceptibility and perceived severity (Liang and Xue 2010). In this study, perceived susceptibility refers to developers' assessment of the likelihood of a data breach occurring via their software. For instance, a developer of a scientific calculator application might not perceive a threat of identity theft, given that the application does not handle personal data. Thus, developers perceive a threat when they recognise the applications' susceptibility to it. Perceived severity indicates the extent to which developers believe a data breach could have a negatively impact organisations (e.g., regulatory fines) or software end-users (e.g., mental distress) (Boteju et al. 2024). However, unlike the TTAT, perceived severity alone does not influence threat perception in developers; it is only considered once they acknowledge susceptibility. For example, knowing the consequences (severity) of GDPR non-compliance does not support perceiving a threat if an application is not handling personal data (Hargitai et al. 2018). Thus, we assert:

H3a: Perceived susceptibility positively affects perceived threat.

H3b: Perceived susceptibility moderated by perceived severity positively impacts perceived threat.

**Safeguard Expectations**

The safeguard expectation component includes three motivational factors: safeguard effectiveness, perceived benefits and affordable costs. Safeguard effectiveness is the ability of a specific safeguard to mitigate identified threats successfully (Liang and Xue 2010), perceived benefits are the additional advantages developers believe they receive by utilising PETs (Boteju et al. 2024), and affordable costs are trade-offs of using PETs that do not outweigh their benefits (Liang and Xue 2010).

IT users are motivated to use a safeguard against a perceived threat if that safeguard can successfully counter the identified threat (Liang and Xue 2010; Tanner et al. 1991). Similarly, developers are motivated to use PETs only if those techniques can mitigate identified threats to personal data (Boteju et al. 2024). Therefore, we assert:

H4: Safeguard effectiveness positively motivates developers to integrate PETs into software.

Software developers are more driven to adopt PETs due to additional benefits that arise with their data protection capabilities (Boteju et al. 2024). PETs allow using personal data for secondary purposes without violating regulations (Sanderson et al. 2023). For example, as GDPR restricts using personal data beyond its original purpose, data cannot be reused for further analysis without the users' explicit consent (Commission 2016). In such cases, developers can use PETs to anonymise the datasets, removing the link to the data owners and, therefore, exempting them from GDPR governance (Sanderson et al. 2023).

H5: Perceived benefits of PETs positively motivate developers to integrate PETs into software.

Safeguard costs are expenses, such as time, money, manpower or utility that are traded off to use the safeguards (Liang and Xue 2010). When IT users incur safeguard costs that surpass the benefits, they are demotivated to use those safeguards (Liang and Xue 2010). Similarly, software developers are reluctant to integrate PETs into software, if there are no available resources to cover the attached costs or if the PETs compromise efficiency (Boteju et al. 2024). For example, developers may avoid using differential privacy on sensor data (e.g., GPS locations) since the added noise further increases their natural inaccuracies (decreased utility) (Munilla Garrido et al. 2023). Thus, we hypothesise:

H6: Affordable costs of PETs positively motivate developers to integrate PETs into software.

According to the TTAT, safeguards effectiveness and perceived threat have a negative interaction effect on avoidance motivation (Liang and Xue 2010). For example, when a perceived threat is higher, IT users are motivated to follow emotion-focused coping rather than relying on safeguards. And users may feel less urgency to avoid perceived threats if they already have effective safeguards in place. However, in the context of developers, we argue otherwise. The higher the safeguard effectiveness is, the stronger the relationship between perceived threat and motivation is (Boteju et al. 2024). For example, assume there are two PETs to resolve the threat of unauthorised location tracking, where the first PET has a higher effectiveness than the other. Once the developers perceive unauthorised location tracking, they are more motivated to use the first PET. We extend the same argument for the interaction between perceived threats and safeguard benefits or affordable costs.

H7: The interaction of perceived threats and safeguard expectations positively impacts the motivation to integrate PETs into software.

**Threat Avoidance Capability**

Even if a developer recognises a threat and selects a PET within their expectations, they still need to be capable of integrating the PET (Boteju et al. 2024). We define this inner control as threat avoidance capability consists of two motivational factors: self-efficacy and competence. In this study, self-efficacy denotes the developers' confidence in their ability to integrate PETs into software, and competence refers to their ability to develop PETs integrated software effectively.

Bandura's self-efficacy theory of motivation claims that a person's self-efficacy determines their behaviour (Bandura 1977). Empirical evidence shows that developers with low self-efficacy in integrating PETs are less likely to use PETs (Boteju et al. 2024). For instance, developers who lack confidence could hesitate to implement PETs, fearing potential blame for incorrect implementation (Munilla Garrido et al. 2023). Thus, self-efficacy empowers developers to integrate PETs with confidence. In addition, Bandura (1977) claims that competence (mastery experience) is a strong source to strengthen self-efficacy. Thus, we hypothesise:

H8: Self-efficacy positively motivates developers to integrate PETs into software.

H8a: Competence increases self-efficacy in integrating PETs into software.

The relationship between competence and motivation is absent in TTAT (Liang and Xue 2010). In this context, developers' lack of competence demotivates them to integrate PETs into software (Boteju et al. 2024) Developers who lack the competency in integrating PETs either refrain from integrating them into software or integrate them ineffectively, giving them a sense of burden (Boteju et al. 2024; Sarathy et al. 2023; Senarath and Arachchilage 2018b). Thus, we argue:

H9: Competence motivates developers to integrate PETs into software.

**Favourable Social Norms**

People are more likely to behave in a particular manner if they believe others approve of that behaviour (Ajzen 1991). Software organisations include stakeholders, such as customers, Chief Privacy Officer (CPO), project managers and quality analysts, each bringing unique perspectives and expertise to the decision-making process (Agrawal et al. 2021; Shilton et al. 2020). These stakeholders directly or indirectly influence developers' perceptions, including those regarding data protection and PETs (Boteju et al. 2024). For example, if higher management does not approve or enforce the decision to integrate PETs into software, developers are demotivated to utilise PETs (Iwaya et al. 2023). Therefore, the more the others value PETs, the more likely developers feel motivated to incorporate PETs into software. "PETs-101" refers to this social pressure as "favourable social norms".

H10: Favourable social norms positively motivate developers to integrate PETs into software.

## *Game Application Scenario*

This section introduces an example game application (Figure 2) that translates the discussed motivational factors into gameplay (assuming the hypotheses hold). The game achieves the motivational factors by simulating threats to personal data, providing knowledge of PETs and simulating privacy levers (Figure 1). Following a storytelling approach, the game allows players to make data protection decisions for a given software application using PETs. The game starts by describing the data protection requirements of the application. The next step is to **simulate the threats to personal data**. Alhazmi and Arachchilage (2023) illustrate scenarios of data breaches caused by poorly developed software. The example game attempts threat simulation through a video call with a Chief Privacy Officer (CPO) (Figure 2a) followed by quizzes. The CPO explains the regulatory and financial consequences (perceived severity) of breaching personal data. Then, the game provides a quiz that asks the players to select the personal data items associated with the application (perceived susceptibility). Next, in Figure 2b, the game tests whether the players recognise the possible threats to the identified personal data (perceived threat). In combination, three efforts adhere to the threat perception in Figure 1.

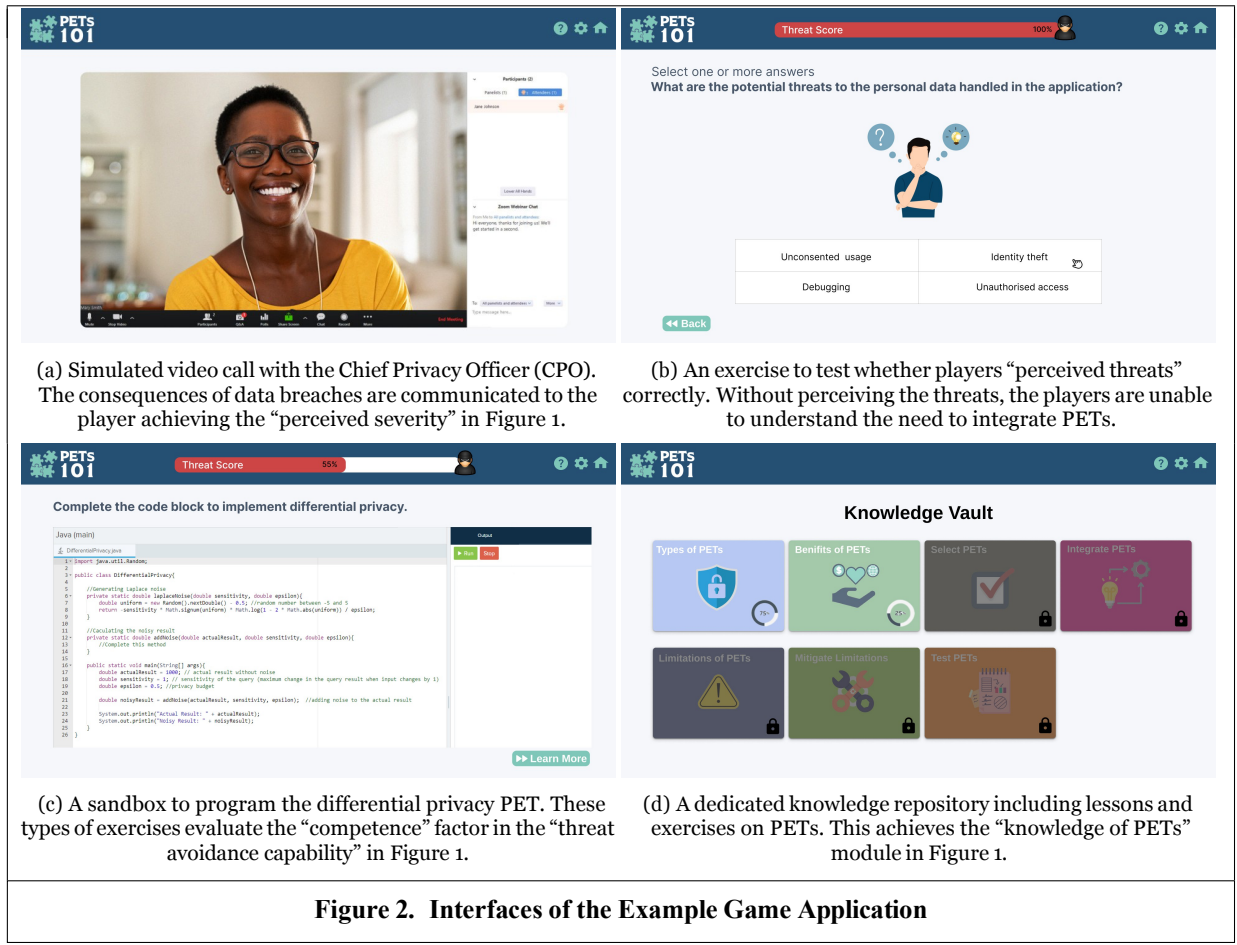

(a) Simulated video call with the Chief Privacy Officer (CPO). The consequences of data breaches are communicated to the player achieving the "perceived severity" in Figure 1.

(b) An exercise to test whether players "perceived threats" correctly. Without perceiving the threats, the players are unable to understand the need to integrate PETs.

(c) A sandbox to program the differential privacy PET. These types of exercises evaluate the "competence" factor in the "threat avoidance capability" in Figure 1.

(d) A dedicated knowledge repository including lessons and exercises on PETs. This achieves the "knowledge of PETs" module in Figure 1.

**Figure 2. Interfaces of the Example Game Application**

Then, the game initiates a threat score and displays it to the players (Figure 2b). The threat score reduces each time the players takes steps to mitigate the identified threats using PETs. For this, the subsequent game scenarios lead players to select suitable PET(s) for the application, integrate them and evaluate the integration. A similar exercise to Figure 2b can be given to check whether players can choose suitable PET(s). Next, the players must implement the selected PET(s). For example, Figure 2c shows a sandbox interface where players must complete a code to implement differential privacy. A similar creative task can be given to evaluate the players' knowledge of testing the implemented PETs.

The players can pause the main game and access the Knowledge Vault (Figure 2d), a dedicated module to **provide knowledge of PETs**. Alhazmi and Arachchilage (2023) also provided knowledge through informative videos to influence self-efficacy. However, in the example game players can achieve all motivational factors grouped under the threat avoidance capability and safeguard expectations through the learnt knowledge. The Vault employs the backward design strategy (McTighe and Wiggins 1999), where learner outcomes are defined first, followed by selecting suitable teaching content. The learner outcomes are structured using the six learning levels of the revised Bloom's Taxonomy (remember, understand, apply, analyse, evaluate and create), which provides a progressive learning experience, improving learners from beginners to experts (Anderson and Krathwohl 2001). A more detailed usage of this taxonomy in PETs related education can be seen in Klymenko et al. (2023). To satisfy the learning outcomes, the Vault provides theoretical and practical knowledge of PETs through lessons and follow-up exercises. For example, the lesson "Types of PETs" includes content on different PETs, their mathematical foundations, and applicable scenarios followed by exercises to measure the learnt knowledge. Each lesson (except the first) is unlocked based on the players' progress in the previous lesson, ensuring satisfactory knowledge acquisition.

Finally, the game simulates privacy levers to achieve favourable social norms. Shilton et al. (2020) simulated this social pressure by simulating an email from the head of the legal department and directing players to developer forums. In this game, the conversation with the CPO (Figure 2a) sets the stage for the players to grasp PETs as socially favoured technologies for data protection. Since the CPO is a stakeholder in the company, the conversation will convey that the corporate culture values data protection. As the game progresses, players will encounter PETs and their ability to achieve data protection. With this newfound understanding, players can naturally link PETs as a favourable social norm to protect personal data.

## Future Work and Potential Contributions

In future, we plan to validate the hypotheses of "PETs-101" framework, implement a game-based learning intervention based on the validated "PETs-101", and evaluate the intervention. First, to verify the proposed hypotheses, a survey study will be conducted with software developers using a Likert questionnaire (Liang and Xue 2010). The responses will be evaluated using the statistical methods suggested by Liang and Xue (2010). We expect to select a diverse participant cohort for this study, considering geographical regions, experience, domain specialisation, gender, and proficiency in PETs. We expect such diversity would generalise the study results, making these results applicable to a broader software developer community (Iwaya et al. 2023).

Next, a game prototype will be implemented upon the validated framework as a web-based approach. A web-based intervention eases the availability and accessibility during user studies. Although the proposed framework applies to any PET, as a proof-of-concept, we will design the game for three PETs from three categories: differential privacy (provable models), k-anonymity (probable models), and pseudonymisation (naive models) (Kaafar et al. n.d.) offering a balanced perspective on different categories of PETs. The prototype's usability will be enhanced through iterative user studies conducted with software developers, incorporating feedback and insights at each iteration (Shilton et al. 2020). The final study, which includes a pre-questionnaire, gameplay, and exit interviews, will be conducted with a disjoint cohort to assess assess the impact of the game intervention. The pre-questionnaire will assess developers' current stance along the framework components using a Likert questionnaire and an exercise to apply their current data protection knowledge. The gameplay will be conducted as a think-aloud study, where the participants will explain their thoughts while navigating the game (Alhazmi and Arachchilage 2023). The exit interviews will include the same pre-questionnaire, pre-defined and ad-hoc (based on the given responses) open-ended questions, and a chance to improve the answer to the exercise completed during the pre-questionnaire. The answers to the open-ended questions will be qualitatively evaluated using the Reflexive Thematic Analysis (Braun and Clarke 2021). Participants for future studies will be identified using the methods from (Klymenko et al. 2022). We will use voluntary recruitment, which is also accomplished in previous studies (Agrawal et al. 2021; Alhazmi and Arachchilage 2023; Dwork et al. 2019; Klymenko et al. 2022).

While dedicated research efforts on PETs are crucial to enhance their capabilities, it is equally vital to explore how to put those efforts into practice. For this, future work can employ our framework to develop different educational interventions focused on PETs (e.g., undergraduate course modules). Educators can use the framework to teach computer science students (i.e., novice developers) about using PETs to achieve data protection technically. In addition, software organisations can adapt it to improve existing privacy training and awareness programs or to introduce new training programs. Such programs also lead developers in conducting standardised risk management processes. This contribution is possible since "PETs-101" aligns with the components of ISO 31000:2018 Risk Management standard (ISO 2018b), such as risk source (susceptibility), consequence (severity), event (threat), involvement of stakeholders (favourable social norms) and risk treatment (selecting PETs that satisfy safeguard expectations and integrating PETs due to threat avoidance capability). Further, researchers can extend our work, investigating more into PETs based education for developers. Ultimately, software developers can improve their privacy-aware software development practices through "PETs-101", enhancing data privacy protection within the digital landscape.

Teaching Software Developers to Integrate PETs into Software